\documentstyle[prl,aps,epsf,multicol]{revtex}
\begin{document}
\sloppy
\draft
\title
{Dynamo mechanism: Effects of correlations and viscosities}
\author
{Abhik Basu\cite{byemail}}
\address{Abteilung Theorie, Hahn-Meitner-Institut, Glienicker Strasse
  100, D-14109 Berlin, Germany,\\and\\
Poornaprajna Institute of Scientific Research,
Bangalore, India.}
\maketitle
\begin{abstract}
We analyze the effects of the background velocity and the initial magnetic
field correlations, and viscosities
on the turbulent dynamo and the $\alpha$-effect. We calculate
the $\alpha$-coefficients for arbitrary magnetic and fluid viscosities, background
velocity and the initial magnetic
field correlations.
We explicitly demonstrate that the general features of the
initial growth and late-time saturation of the magnetic fields 
due to the non-linear feedback are qualitatively independent of these
correlations. We also examine the hydrodynamic limit of the
magnetic field growth in a renormalization group framework and discuss the
possibilities of suppression of the dynamo growth below a critical rotation. 
We demonstrate that for Kolmogorov- (K41)
type of spectra the {\em Ekman number } $M \stackrel{>}{\sim}1/2$ 
for dynamo growth to occur.

\end{abstract}

\pacs{PACS no:47.65.+a,91.25.Cw}
\section{Introduction}
Magnetic fields are ubiquitous. All astrophysical objects are known to have
magnetic fields of different magnitudes,e.g., 1 gauss at the stellar scale
to $10^{-6}$ gauss at the galactic scale \cite{moff}. The origin of such
fields ({\em primordial field}) is not very clear - there are several 
competing theories which attempt to describe this \cite{kron}. However,
a finite magnetic field in any physical system undergoes a temporal 
decay due to the
finite conductivity of the medium. So, for steady magnetic fields 
to occur in astrophysical bodies, there has to be a mechanism of 
regeneration of the
magnetic fields, which takes place due to the dynamo process
\cite{moff,abhik}. Most astrophysical bodies are thought to have
{\em fast} dynamo operating within themselves (there are exception to this, e.g., the
Moon, Venus and Mars in our solar system) resulting into 
exponential growth of the magnetic fields. This mechanism requires 
a turbulent velocity background \cite{moff} [though non-turbulent velocity 
fields too can make a seed (initial) magnetic field to grow (for 
details see \cite{abhik}), we will not consider such cases here].
Since the dynamo equation, in the linear approximation
(see below) gives unbounded exponentially growing solutions for the long
wavelength (large scale) part of the magnetic fields, it is linearly
unstable in the low wavenumber limit.  
However, one does not see a perpetual growth of  magnetic fields in the core
of the earth or in the sun. For example, geomagnetic fields ($\sim$ 1 gauss)
are known to be stable for about $10^6$ years \cite{moff}.
Thus, the physically realisable
solutions of the dynamo equations cannot be unstable in the long time limit.  
It is now believed that
the non-linear feedback due the Lorentz force term in the Navier-Stokes
equation is responsible for the saturation of the magnetic field growth (see,
e.g., \cite{moff}).

The study of this problem has already been the subject of previous work by 
many groups. For example
Pouquet, Frisch and L\'{e}orat \cite{po}
studied the connections between the dynamo process and the
inverse cascade of magnetic and kinetic energies within a eddy damped quasi-normal
Markovian
approximation. Moffatt \cite{moff1} has examined the back reactions due to the
Lorentz force for magnetic Prandtl number $P_m\gg 1$ by
linearising the equations of motion of three-dimensional ($3d$)
magnetohydrodynamics (MHD). Vainshtein and Cattaneo
\cite{vain} discussed several nonlinear restrictions on the generations of
magnetic fields. Field {\em et al} \cite{field} discussed nonlinear
$\alpha$-effects within a two-scale approach. Rogachevskii and Kleeorin
\cite{roga} studied the effects of an anisotropic background turbulence on the
dynamo process. Brandenburg examined non-linear $\alpha$-effects in numerical
simulation of helical MHD turbulence \cite{brand}. In particular, he examined the
dependences of dynamo growth and the saturation field on the magnetic Prandtl
number $P_m$ (the ratio of the magnetic- to the kinetic- viscosities). 
Bhattacharjee and Yuan \cite{bhatta} studied the problem in a two-scale approach by
linearising the equations of motion. 

Dynamo mechanism has two competing processes at work: amplification of the
magnetic field by the dynamo process and ohmic dissipation due to finite
resistivity of the medium concerned. Which one among these two effects 
will dominate depends on the case in study. In some
specific models, however, one can analyze this completely. A good example of
such models is the Kraichnan-Kazantzev dynamo \cite{kraich,kaz} where the
velicity field is assumed to be Gaussian-distributed, delta-correlated in time
and the magnetic field is governed by the Induction equation \cite{jac}. In
this model the statistics of the velocity field is taken to be parity invariant
so that the $\alpha$-effect is ruled out. The main results from this model
include i)the existence of dynamo in the infinite magnetic Reynolds number limit 
for a
particular choice of the variance of the velocity distribution \cite{var} and
ii)the existence of a critical magnetic Reynolds number only 
above which dynamo growth
is possible \cite{vin}. However, not
much is known about this when invariance due to parity is broken and when the 
velocity field
is not temporally delta-correlated. In a recent simulations \cite{brand1} 
the authors found, in a model simulation for
the solar convection zone, a monotonic increase of the horizontal
$\alpha$-effect with rotation. Kida {\em et al} showed, in numerical simulations,
that unless magnetic hyperviscosity is less than a critical value, magnetic
fields did not grow \cite{kida}, confirming the existence of a critical
magnetic Reynolds number ($R_m$).
 
Our studies generalize the existing results.
In this paper we use a minimal model of $\alpha$-effect (see below) to 
study dynamo with $\alpha$-effect to calculate the $\alpha$
coefficient for arbitrary correlations and viscosities, and ask the 
following questions:
\begin{enumerate}
\item Do the turbulent dynamo growth and the saturation processes require any turbulent
background? Or do they function with arbitrary parity-breaking and 
fluctuating velocity and initial magnetic field correlations?
{\footnote{By a turbulent background we do not mean any kind of
fluctuating state but
a fluctuating state with Kolmogorov (K41) spectra $\propto k^{-5/3}$ 
for the kinetic and magnetic energies and
cascades of appropriate quantities; if there is no mean magnetic field then the
energy spectra is expected to be K41-type - see Ref.\cite{abprl}.} }. 
\item What is the {\em hydrodynamic limit} (long wavelength limit) of the
dynamo problem? By this we ask how the magnetic field correlations scale
in the infra red limit during the initial-growth regime.
\item Can arbitrarily large magnetic viscosity prevent dynamo growth?
In other words, is there a critical magnetic Reynolds number $R_m$ above which the
dynamo growth sets in?
\end{enumerate}
To study the above mentioned questions we employ a 
diagrammatic perturbation
theory, which has been highly successful in the contexts of critical dynamics
\cite{halparin}, driven systems \cite{fns}, etc. This
can be easily extended to
higher orders in perturbation expansion and  is very suitable for
handling continuous kinetic and magnetic spectra, unlike the two-scale approximation.
This was first used to study
stationary, homogeneous and isotropic MHD in Ref.\cite{lee}. We use
this method to study non-stationary statistical states (dynamo growth) which facilitates
studies on the hydrodynamic limit of the dynamo problem in a renormalisation group
framework. We use diagrammatic perturbation theory to calculate expressions for
the $\alpha$ coefficiants for arbitrary background velocity and initial magnetic
field correlations and magnetic Prandtl number $P_m$ for both the 
early growth and the late time saturation. With our expressions for
$\alpha$ we examine the three issues mentioned above.

We investigate these for arbitrary correlations and magnetic Prandtl number $P_m$
with no approximations other than the existence a perturbation theory. Our
principal results are:
\begin{itemize}
\item We calculate the $\alpha$-coefficients for arbitrary correlations and
viscosities.
\item We examine the hydrodynamic limit in the kinematic regime and predict the
existence of a critical $R_m$ or rotation above which dynamo growth will occur
for certain correlations with infra red singularity.
\end{itemize}
In our all our studies, we do not assume any variance for the velocity field.
Instead, we use the Navier-Stokes equation to describe the dynamics of the
velocity field. This allows us to use a renormalisation group framework to study
the hydrodynamic limit.

The first question that we investigate 
is phenomenologically very important because different systems
may have different velocity and initial magnetic field spectra. 
Therefore, it is important to 
understand the dependence of the dynamo on these spectra.
We explicitly demonstrate that the nonlinear feedback of the magnetic fields on
the velocity fields in the form of the Lorentz force stabilises the
growth for arbitrary velocity and initial magnetic field
correlations. This demonstrates that the basic features of the dynamo mechanism
are qualitatively independent of the velocity and magnetic field spectra and,
essentially, are properties of the $3d$MHD equations.
Details (e.g., the values of the $\alpha$-coefficients)
of course, depend upon the actual forms of the spectra.
Our renormalization group
 analysis indicates that dynamo growth takes place only if the Ekman number
$M\lesssim 1/2$ (for a given $R_m$) when the velocity and the initial
magnetic field spectra are sufficiently singular in the long wavelength limit. 
The structure of this paper is as follows:
In Sec.\ref{dyn} we discuss the general dynamo mechanism
within the standard linear approximation for arbitrary velocity and initial
magnetic field correlations and viscosities. In Sec.\ref{halt} we
show that beyond the linear approximation 
non-linear effects lead to the
eventual saturation of magnetic field growth for arbitrary 
background kinetic
energy and initial magnetic energy spectra, and viscosities. 
We elucidate how different  background
kinetic energy and initial magnetic energy spectra affect the values of the
$\alpha$-coefficients. In Sec.\ref{renor} we analyze the initial dynamo growth
in a renormalization group framework. We show that for sufficiently singular
velocity and magnetic field spectra the Ekman number must be $\lesssim$ 1/2
for the magnetic fields to grow. For velocity and magnetic field
spectra which go to zero in the long wavelength limit there are no such
restrictions.
In Sec.\ref{summ} we present our conclusions. 

\section{Dynamo growth: the linear approximation}
\label{dyn}
In the kinematic approximation \cite{moff,abhik1}, i.e., in the early-time regime,
when the
magnetic energy
is much smaller than the kinetic energy ($\int u^2 d^3r >> \int b^2
d^3 r$, where ${\bf u(r},t)$ and ${\bf b(r},t)$ are the velocity and magnetic
fields respectively) the Lorentz force term of the Navier Stokes 
equation is neglected. In that weak magnetic field limit, which is 
reasonable at an early time, the time evolution problem for the magnetic
fields is a linear problem as the Induction equation \cite{jac}
is linear in magnetic fields $\bf b$:
\begin{equation}
{\partial {\bf b}\over \partial t}=\nabla\times ({\bf u\times b})+\mu
\nabla^2 {\bf b},
\label{indeq}
\end{equation}
where $\mu$ is the magnetic viscosity.  The velocity field is governed by the
Navier-Stokes equation \cite{land} (in the absence of the Lorentz force)
\begin{equation}
{\partial{\bf u}\over\partial t}+{\bf u.\nabla u}=-{\nabla p\over\rho}
+\nu\nabla^2 {\bf u}+{\bf f}.
\end{equation}
Here $\nu$ is the fluid viscosity,  $\bf f$ an external forcing function,
$p$ the pressure and $\rho$ the density of the fluid.
We take $\bf f$ to be a zero mean, Gaussian stochastic force with a
specified variance (see below).

In a two-scale \cite{moff} approach one can then write an {\em effective} 
equation for ${\bf B}$, the long-wavelength part of the magnetic fields 
\cite{moff}:
\begin{equation}
{\partial {\bf B}\over \partial t}=\nabla\times ({\bf U\times B})+
\nabla\times {\bf E}+\mu\nabla^2 {\bf B},
\end{equation}
where the {\em Electromotive force} $\bf E=\langle u\times b\rangle$. 
$\bf U$ is the large scale component of the velocity field $\bf u$. An 
{\em Operator Product Expansion} (OPE) is shown to hold \cite{abhik1} 
which provides a gradient expansion in terms of $\bf B$ for the product  
$\bf E=\langle u\times b\rangle$ \cite{moff}
\begin{equation}
E_i=\alpha_{ij}B_{j}+\beta_{ijk}{\partial B_{j}\over \partial x_k}+... .
\label{ope}
\end{equation}
For homogenous and isotropic flows ($\alpha_{ij}=\alpha\delta_{ij}$) 
Eq.(\ref{ope}) gives,
\begin{equation}
{\partial {\bf B}\over\partial t}=\nabla\times ({\bf U\times B})
+\alpha\nabla\times {\bf B}+\mu\nabla^2{\bf B},
\label{dyna}
\end{equation}
which is the standard turbulent dynamo equation. Here $\mu$ now is the
{\em effective} magnetic viscosity which includes both the microscopic magnetic
viscosity and the turbulent diffusion,
represented by $\beta_{ijk}$ in Eq.(\ref{ope}). $\alpha$  depends
upon the statistics of the velocity field (or, equivalently, the correlations of
$\bf f$). Retaining only the $\alpha$
-term and dropping all others from the RHS of Eq.(\ref{dyna}), the
equations for the cartesian components of $\bf B$ become (we neglect the
dissipative terms proportional to $k^2$ as we are interested only in the long
wavelength properties)\\
\[
{d\over dt}
\pmatrix{
B_x({\bf k},t) \cr B_y({\bf k},t) \cr B_z({\bf k},t)}
=i\alpha\pmatrix{
0 & -k_z & k_y \cr
k_z & 0 & -k_x \cr
-k_y & k_x & 0}
\pmatrix{
B_x({\bf k},t) \cr B_y({\bf k},t) \cr B_z({\bf k},t)}.
\]
The eigenvalues of the matrix is {$\lambda=\pm ik,\,0$}. Thus depending on the
sign of the product $\alpha k$, one mode grows and the other decays. 
The third mode is unphysical,
because the corresponding eigenfunction is proportional to $\bf{k}$ and
hence in conflict with $\nabla\cdot\bf {B}=0$.
Since growth rate is proportional to $|k|$ and dissipation is
proportional to $k^2$, large scale fields continue to grow leading to 
long wavelength
instability. Thus in the long time limit effectively only the growing mode
remains.  
Growth rate $\alpha$ is a pseudo-scalar quantity, i.e., under
parity transformation $\bf r\rightarrow -r$, $\alpha\rightarrow -\alpha$ 
\cite{moff,abhik1}. Since $\alpha$ depends upon the statistical properties of
the velocity field, its statistics  should not be parity invariant.
This can happen in a rotating frame, where the angular
velocity explicitly breaks reflection invariance.

\section{Formulation of the dynamo problem in a rotating frame}
\label{rot}
The Navier-Stokes (NS) (including the Lorentz force) 
and the Induction equation in an inertial frame in $({\bf k}, t)$ space take the
form
\begin{equation}
{\partial u_i({\bf k},t)\over\partial t}+{1\over 2}P_{ijp}({\bf k})
\sum_q u_j({\bf q},t)
u_p({\bf k-q},t)={1\over 2}P_{ijp}({\bf k})\sum_q b_j({\bf q},t)
b_p({\bf k-q},t)- \nu k^2 u_i +f_i({\bf k},t),\\
\label{nsk}
\end{equation}
and
\begin{equation}
{\partial b_i({\bf k},t)\over\partial t}=\tilde{P}_{ijp}({\bf k})
\sum_q u_j({\bf q},t) b_p({\bf k-q},t)-\mu k^2 b_i.
\label{indk}
\end{equation}
Here, $u_i({\bf k},t)$ and $b_i({\bf k},t)$ are the fourier transforms of
$u_i({\bf r},t)$ and $b_i({\bf r},t)$ respectively, $P_{ijp}({\bf k})=P_{ij}
({\bf k})k_p+P_{ip}({\bf k})k_j,\,\tilde{P}_{ijp}({\bf k})=P_{ij}({\bf k})k_p
-P_{ip}({\bf k})k_j,\,P_{ij}$ is the projection operator, which appears due to
the divergence-free conditions on the velocity and magnetic fields (we consider
incompressible fluid for simplicity). 
Equations (\ref{nsk}) and (\ref{indk}) have to be supplemented by appropriate
correlations of $f_i$ and initial conditions on $b_i$. We choose $f_i({\bf
k},t)$ and $b_i({\bf k},t=0)$ to have zero mean and to be
Gaussian distributed with the
following variances:
\begin{equation}
\langle f_i({\bf k},t)f_j(-{\bf k},0)\rangle=2P_{ij}D_1(k)\delta(t),\\
\label{varf}
\end{equation}
\begin{equation}
\langle b_i({\bf k},t=0)b_j(-{\bf k},t=0)\rangle=2P_{ij}D_2(k),
\label{varb}
\end{equation}
where
$D_1$ and $D_2$ are some functions of $k$ (to be specified later).

In a rotating frame with a rotation velocity ${\bf \Omega}=\Omega\bf \hat{z}$ 
the Eq.(\ref{nsk}) takes the form
\begin{equation}
{\partial u_i({\bf k},t)\over\partial t}+2({\bf \Omega\times u})_i+
{1\over 2}P_{ijp}({\bf k}) \sum_q u_j({\bf q},t)
u_p({\bf k-q},t)={1\over 2}P_{ijp}({\bf k})\sum_q b_j({\bf q},t)
b_p({\bf k-q},t)+ \nu\nabla^2 u_i +f_i({\bf k},t),
\label{nskr}
\end{equation}
whereas Eq.(\ref{indk}) has the same form in the rotating frame.
$\bf \Omega\times u$ is the coriolis force. The centrifugal force
$\bf \Omega\times(\Omega\times r)$ is a part of the {\em effective
pressure}=$p+{1\over 2}|{\bf \Omega\times r}|^2$ which does not contribute to
the dynamics of incompressible flows.
The bare propagator $G_u$ (obtained from the linearized  version of
Eq.(\ref{nskr}))
of $u_i$ \\
\[G_u=\pmatrix{{{i\omega+\nu k^2}\over (i\omega +\nu k^2)^2+4\Omega^2}& 
-{2\Omega\over (i\omega+\nu k^2)^2+4\Omega^2} 
& 0 \cr
{2\Omega\over (i\omega+\nu k^2)^2+4\Omega^2} & {{i\omega +\nu k^2} \over
(i\omega +\nu k^2)^2+4\Omega^2}& 0 \cr
0 & 0 & {1\over i\omega+\nu k^2}} 
\]
\\
such that $\bf u= G_u\,f$ where $\bf u$ is the column vector
\[{\bf u}=\pmatrix{u_x\cr u_y\cr u_z}.
\]
\\
One can verify that with the form of the bare propagator given above, an
odd-parity part in the velocity auto-correlator $\langle u_i({\bf k}, t)u_j({\bf
-k}, 0)\rangle$ appears which is proprotional to the rotation $\Omega$. 
Notice that $G^{zz}_{u}$
is different from $G^{xx,yy}_{u}$ - this is just the consequence of the
fact that $\Omega$ distinguishes the $z$-direction as a preferred direction in
space, making the
system anisotropic. However for frequencies $\omega >>\Omega$ or length scales
$k^z>>\Omega$ (here $z$ is the dynamical exponent)
isotropy is restored. In that regime, to $O(\Omega)$ the role of
the global rotation is to introduce a non-zero odd-parity part in
$\langle u_iu_j\rangle$ proportional to $\Omega$.
This can be also seen by noting that in the inertial frame the
correlation
$\langle u_i({\bf k})u_j({\bf -k})\rangle$ is of the form $P_{ij}({\bf k})A(k)$ [cf. 
Eq.(\ref{varf})] where $A(k)$ is a scalar function of $k$ 
and hence in the rotating frame the correlator is proportional to 
$RP_{ij}R^T$ where $R$'s are appropriate rotation matrices (we have suppressed
the indices). Similarly, initial
magnetic field correlations, given by Eq.(\ref{varb}) transforms accordingly in the
rotating frame. Since rotation matrices act on $\langle u_i({\bf k})u_j({\bf
-k})\rangle$ and Eq.(\ref{varb}) in the same way, magnetic field auto-correlator
$\langle b_i({\bf k},t)b_j({\bf -k},0)\rangle$ has an odd parity part in the
rotating frame with the same sign as the odd parity part in the velocity
correlator.
Thus the effects of rotation can
be {\em modeled} (to the lowest order) 
by introducing parity breaking parts in Eqs.(\ref{varf}) and
(\ref{varb})\cite{moff}
\begin{eqnarray}
\langle f_i({\bf k},t)f_j(-{\bf k},0)\rangle&=&2P_{ij}D_1(k)\delta(t)+
2i\epsilon_{ijp}k_p\tilde{D}_1(k)\delta(t),\nonumber \\
\langle b_i({\bf k},t=0)b_j(-{\bf k},t=0)\rangle&=&2P_{ij}D_2(k)\delta(t)+
2i\epsilon_{ijp}k_p\tilde{D}_2(k),
\label{varbi}
\end{eqnarray}
in conjunction with the Eqs.(\ref{nsk}) and (\ref{indk}), where 
$\epsilon_{ijp}$ is
the totally antisymmetric tensor in $3d$.
This way of modeling rotation effects is, of course, only approximate, but
suffices for our purposes as this explicitly incorporates parity breaking.
One can, however, construct experimental set ups \cite{moff}
which are described correctly by Eqs.(\ref{varbi}). The
parity breaking parts in the noise correlations or initial conditions ensure
that the velocity and the initial magnetic field correlators have non-zero odd
parity parts, as would happen
in a rotating frame. An important dimensionless number is the {\em Ekman number}
$M={\nu L^2\over 2\Omega}$ which can be related to $\tilde{D}_1$ by equating the
parity braking parts of the velocity correlator
 calculated from (linearized) Eq.(\ref{nskr}) and Eq.(\ref{varf}) 
with that from Eqs.
(\ref{nsk}) and (\ref{varbi}). This gives $\tilde{D}_1=2M^{-1}D_1$.
Now, one may ask what is the relative sign between $\tilde{D}_1$ and 
$\tilde{D}_2$? Since the parity breaking parts of the correlators of the velocity
and the magnetic fields have same sign and are proportional to
$\tilde{D}_1 \;{\rm and}\; \tilde{D}_2$ respectively, 
$\tilde{D}_1$ and $\tilde{D}_2$ must have same sign. As already noted,
introduction of parity
breaking terms in the force/initial correlations is well-known in the literature,
we, nevertheless,
give the analysis in details in order to emphasise on the fact that fluid and
magnetic helicities must have the same sign. Furthermore, for a complete 
description of the effects of rotation, in
addition to the coriolis force, a forcing with a preferred direction is also 
required. We, however, do not 
include all these details as introduction of parity-breaking correlations is
sufficient for our purposes. In this sense, this can
be thought of as a {\em reduced} or a {\em minimal model} for dynamo. One may note
that a nonzero kinetic helicity is required for the $\alpha$-effect as the
$\alpha$-coefficient is proportional to the kinetic helocity. Even though a global
rotation explicity breaks the parity invariance of the system under space
reversal, rotation alone is not enough to yield a non-zero helicity. This is
because the helicity is pseudo-scalar and, therefore, can be constructed only out
of an axial vector (here, rotation $\bf \Omega$) and a polar vector. In typical
astrophysical settings, the latter one could be provided by, say, a density
inhomogeneity. Even though this is not contained in Eq. (\ref{nsk}), our minimal
model, nevertheless, produces a finite helicity due to the helical nature of the
forcing function. Thus, our minimal model is able to capture both 
the breakdown of parity
due to the rotation and the generation of helicity due to the rotation and any
other preferred direction.

\subsection{The $\alpha$ in the kinematic approximation: Dependences on 
background velocity and initial magnetic field spectra}

\begin{figure}[htb]
\centerline{
\epsfxsize=8cm
\epsffile{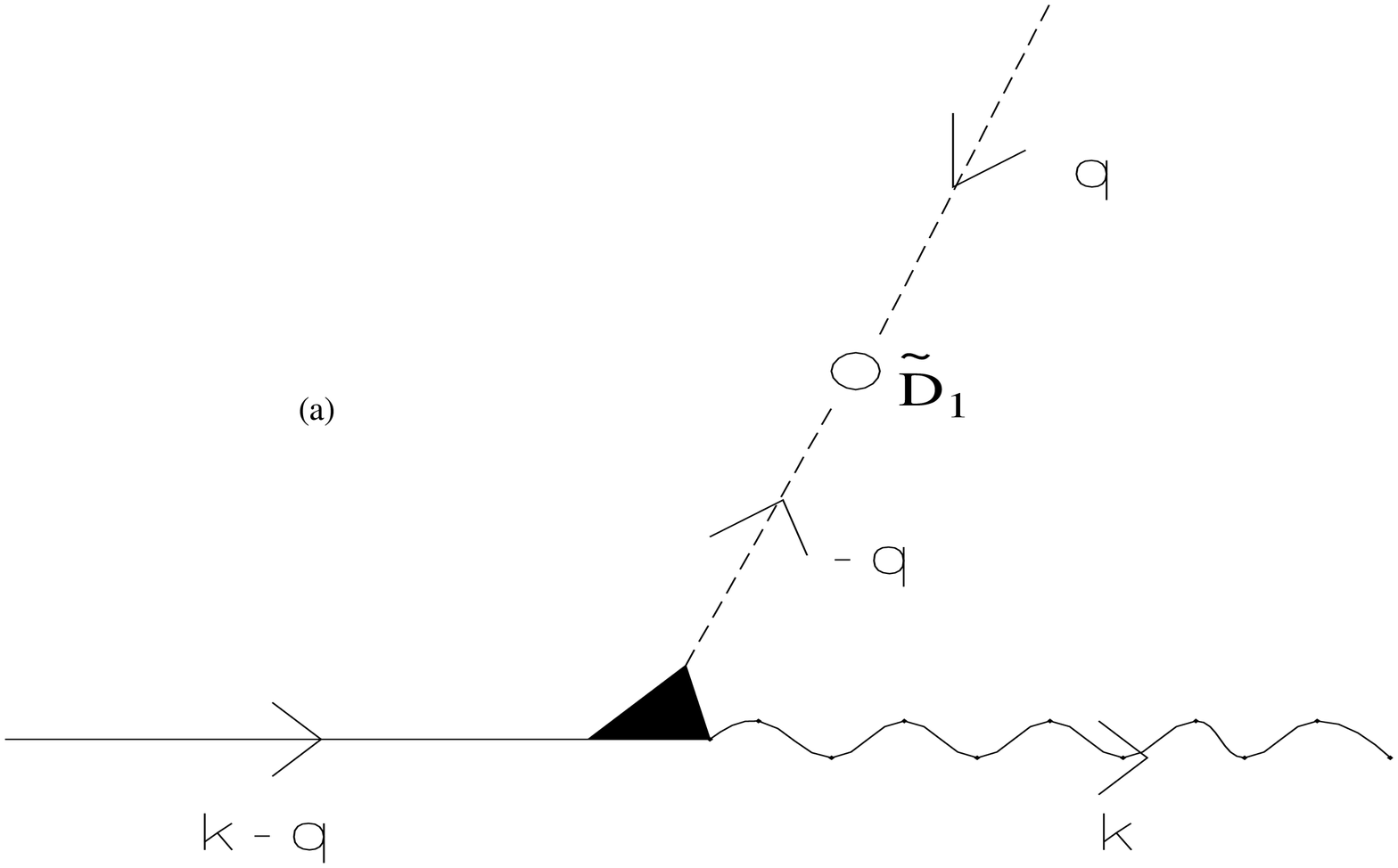}
\hfill
\epsfxsize=8cm
\epsffile{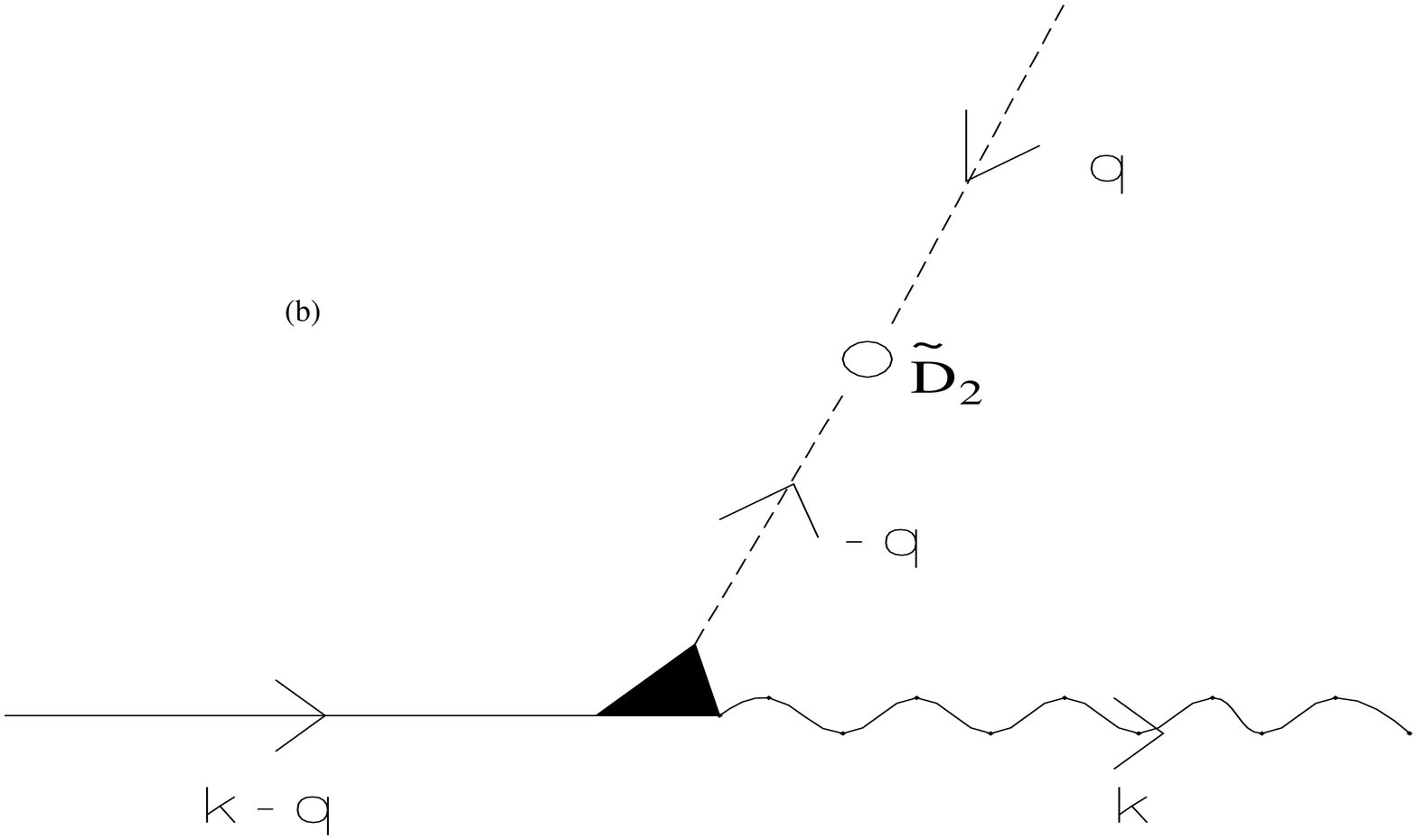}
}
\caption{Tree level diagrams for $<\bf u(q)\times b(k-q)>$. (a)Contribution to
growth term $\alpha_D$: A solid line
indicates a bare magnetic field response function, a broken line indicates
a bare velocity response function, a 'o' joined by two broken lines indicates
a bare velocity correlation function (proportional to $\tilde{D}_1$), 
a wavy line indicates a magnetic field,
a solid triangle indicates a $ub$ vertex.  
(b) Contribution to feedback term $\alpha_F$: A solid line
indicates a bare magnetic field response function, a broken line indicates
a bare velocity response funtion, a 'o' joined by two broken lines indicates
a bare magnetic field correlation function (proportional to $\tilde{D}_2$), 
a wavy line indicates a magnetic field,
a solid triangle indicates a $ub$ vertex.}
\label{diag}
\end{figure}

In the kinematic approximation, which neglects the Lorentz force term of the 
Navier-Stokes equation, the time evolution of the magnetic 
fields follows from the linear Induction Equation (\ref{indeq}). 
We assume, for the convenience of calculations,
that the velocity field ($\bf u$) has reached a statistical steady state.
This is acceptable as long as the loss due to the transfer of kinetic energy to
the magnetic modes by 
the dynamo process is compensated by the external drive. In the 
kinematic (i.e., linear) approximation, we work with the Eqs.(\ref{nsk})
(without the Lorentz force) and (\ref{indk}).
We choose $f_l({\bf k},t)$ to be a zero-mean,
 Gaussian random field with correlations
\begin{equation}
\langle f_l({\bf k},t)f_m({\bf k},0)\rangle=2P_{lm}D_1(k)\delta(t)+
2i\epsilon_{lmn} k_n\tilde{D}_1(k)\delta(t).
\end{equation}
Our initial conditions for the magnetic fields are
\begin{equation}
\langle b_{\alpha}({\bf k},t=0)b_{\beta}(-{\bf k},t=0)\rangle=2P_{\alpha\beta}
D_2(k)+2i\epsilon_{\alpha\beta\gamma}k_{\gamma}\tilde{D}_2(k),
\end{equation}
Since we are interested to investigate the dynamo process with arbitrary
statistics for the velocity and magnetic fields we work with arbitrary $D_1(k),
\tilde{D}_1(k),D_2(k)$ and $\tilde{D}_2(k)$. For K41-type spectra, we require
\cite{yakhot} $D_1(k)=D_1k^{-3},\tilde{D}_1(k)=\tilde{D}_1k^{-4},D_2(k)=D_2
k^{-5/3}$ and $\tilde{D}_2(k)=k^{-8/3}$. These choices ensure that under
spatial rescaling ${\bf x}\rightarrow l{\bf x}$, ${\bf u,b}\rightarrow l^{1/3}
\{ {\bf u,b}\}$ which is the Kolmogorov scaling \cite{yakhot}. 
Note that both the force correlations in the Eq.(\ref{nsk}) and the
initial conditions on Eq.(\ref{indk}) have parts that are parity
breaking, in conformity with our previous discussions. 
We now calculate the $\alpha$-term. We use an iterative perturbative method which
is very similar to and discussed in details in Ref.\cite{fns}. In this method,
terms in each order of the perturbation series can be represented by appropriate
Feynman diagrams \cite{fns}. Even though, for simplicity, 
we confine ourselves to the lowest
order in the perturbation theory (represented by the {\em tree level diagrams}),
which is sufficient for our purposes, higher order calculations represented by
higher order digrams can be  done in a straight forward manner.
Below we give the expression for $\alpha$
in the kinematic approximation (which we call the
`direct' term - responsible for growth) in the lowest order of the perturbation
theory (see Fig.\ref{diag}a):
\begin{equation}
\langle ({\bf u\times b})_{\mu}\rangle_D=\langle
\int_{q,q_1}\epsilon_{\mu\beta\gamma}
u_{\beta}({\bf q},t)b_{\gamma}({\bf k-q},t)\rangle =\langle
\int_q \epsilon_{\alpha
\beta\gamma}u_{\beta}({\bf q},t)\epsilon_{\gamma\delta\lambda}i({\bf k-q})
_{\delta}u_{\eta}({\bf q_1},t_1)b_{\tau}({\bf k-q-q_1},t_1)G_0^b({ k-q},
t-t_1)\rangle
\end{equation}
from which one can read the $\alpha$-term:
\begin{equation}
\alpha_D B_{\alpha}({\bf k},t)=\int_q {i\tilde{D}_1(q) \over \nu q^2}
\epsilon_{\beta\eta\rho}q_{\rho}\epsilon_{\alpha\beta\gamma}(-i)q_{\delta}
b_{\rho}({\bf k},t=0)[{1\over q^2(\nu+\mu)}+{\exp(-2t\nu q^2)\over
q^2(\nu-\mu)}]
\end{equation}
giving
$\alpha_D={2S_3\over 3}{1\over {\nu(\nu+\mu)}}\int_q 2{\tilde{D}(q)\over
(\nu+\mu)q^2}$
for large $t$. 
The suffix $D$ refers to {\em growth} or the {\em direct} term, as opposed to
{\em feedback} which we discuss in the next Sec.\ref{halt}. The growth term is
proportional to $|k|$ and diffusive decay proportional to $k^2$. The angular
brackets represent averaging over the noise and initial-condition ensembles.

\begin{figure}[htb]
\centerline{
\epsfxsize=8cm
\epsffile{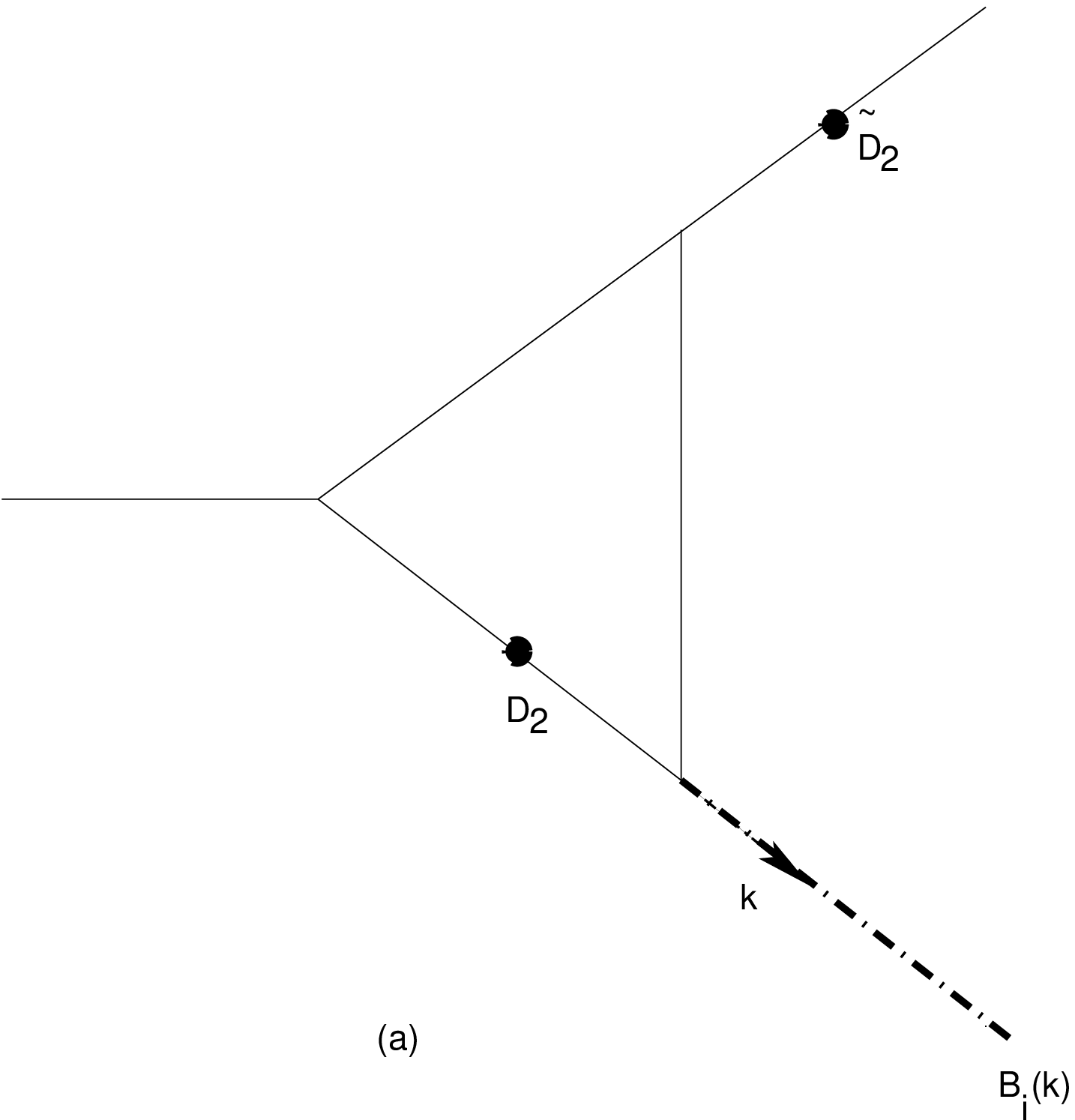}
\hfill
\epsfxsize=8cm
\epsffile{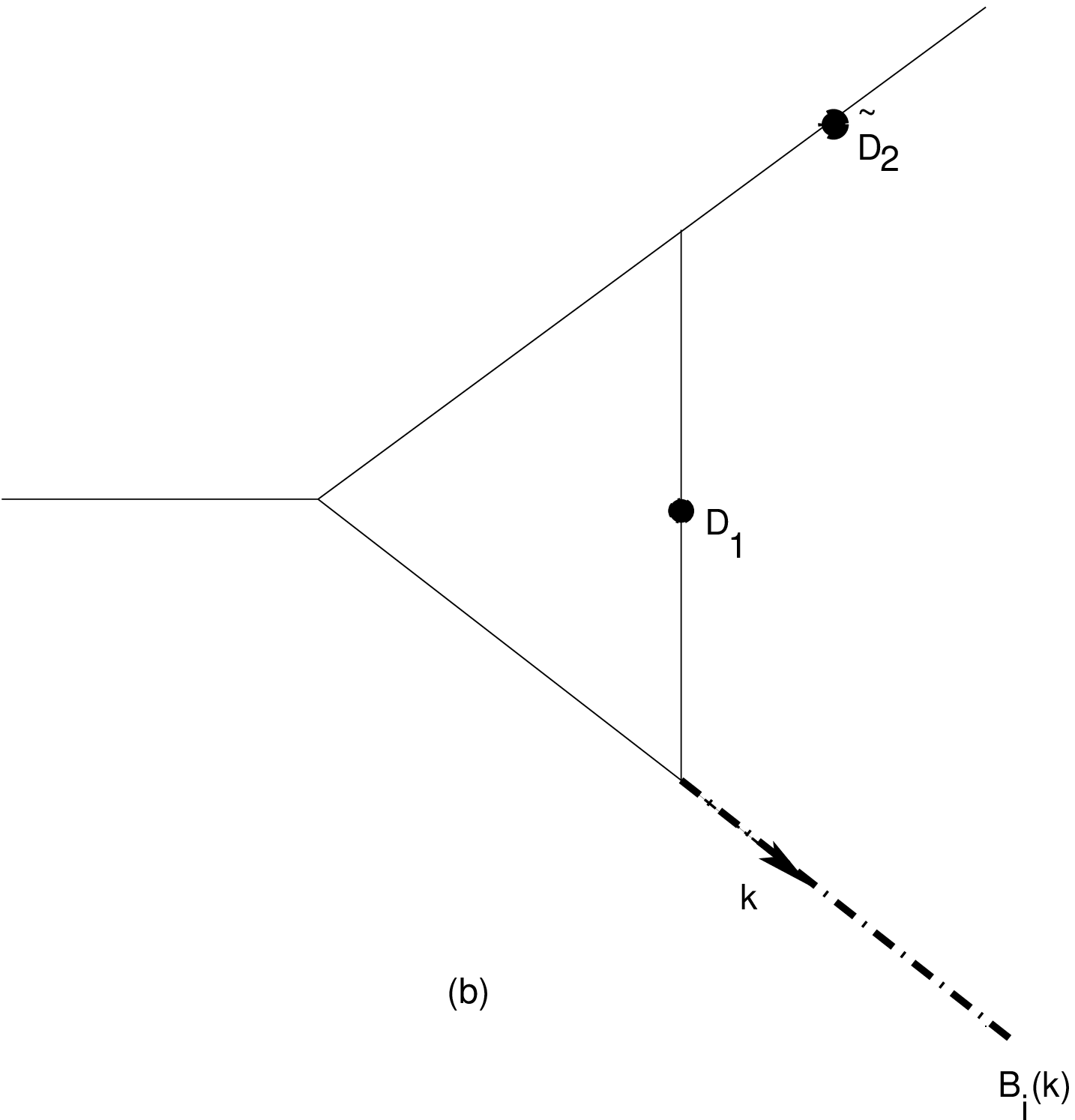}
}
\caption{Two one-loop diagrams contributing to $\alpha_F$. There are total six
diagrams altogether.}
\label{oneloop}
\end{figure}
\subsection{Suppression of growth rate: Nonlinear feedback}
\label{halt}
When the magnetic fields become strong, it is no longer justified to neglect
the feedback of the
magnetic fields in the form of the Lorentz force.
So we need to 
work with the {\em full} Eqs.(\ref{nsk}) and (\ref{indk}). The ideas of
OPE as elucidated in Sec.\ref{dyn} are still valid for the full non-linear
problem. But the value of $\alpha$ is expected to change from its value in the
linear problem.
In presence of the Lorentz force there is an additional contribution to 
$\alpha$ (Fig.1b). To evaluate that,
we follow a diagrammatic perturbation approach similar to that described in
the previous Section. Here also we restrict ourselves to the lowest order only
(i.e., the tree level diagrams) though extension to higher orders is straight
forward. We obtain
\begin{eqnarray}
\langle ({{\bf u\times b}_i})_F\rangle&=&\langle
\int_q \epsilon_{ijp}u_j({\bf q},t)
b_p({\bf k-q},t)\rangle\\=\langle{i\over 2}\epsilon_{ijp}\int_q 
P_{jmn}(q)G_o^u(q,t-t_1)&&
b_m({\bf q_1},t_1)b_n({\bf q-q_1},t_1)G_o^b({k-q},t)b_p({\bf k-q},t=0)\rangle
\end{eqnarray}
which gives ($F$ refers to feedback) 
\begin{equation}
\alpha_F B_i({\bf k},t)=i\epsilon_{ijp}\int_q P_{jmn}(q)e^{2\alpha_D|q|t-2\mu
q^2t}b_n({\bf k},t){-2i\tilde{D}_2(q)\epsilon_{mps}q_s\over 2\alpha_D |q|
-2\mu q^2},
\end{equation}
which, after some simplifications, yields, 
\begin{equation}
\alpha_F(t)={2S_3\over 3}{4\over 15}\int_q {\tilde{D}_2 (q,t) q^2
\over \alpha_D|q|-2\mu q^2},
\end{equation}
where $\tilde{D}_2(q,t)=\exp[2\alpha_D|q|t-2\mu q^2t]\tilde{D}_2(q)$ is
a growing function of time for small wavenumbers. As before, angular brackets
refer to averaging over noise and initial-condition ensembles.
Thus $\alpha_F$ grows in
time.

Since, at any finite time $t$, when the non-linear feedback on the velocity field 
due to the Lorentz force is nolonger negligible, both  $\alpha_D$
and $\alpha_F$ are non-zero and we get
\begin{eqnarray}
\alpha_D&=&-{2S_3\over 3}\int {d^3q\over (2\pi)^3}{\tilde{D}_1(q)\over
{\nu [|(\alpha_D+\alpha_F)q|-(\nu+\mu)q^2]}},\nonumber \\
\alpha_F&=&{2S_3\over 3}{4\over 15}\int {d^3q\over (2\pi)^3}{\tilde{D}_2(q,t)q^2
\over {|(\alpha_D+\alpha_F)q|-2\mu q^2}},
\label{divexp}
\end{eqnarray}
with 
\begin{equation}
\tilde{D}_2(q,t)=\exp [2(\alpha_D-\alpha_F(t))|q|t-2\mu q^2 t]\tilde {D}_2(q).
\label{selfd2}
\end{equation}
Equations (\ref{divexp}) and (\ref{selfd2}) are to be solved self-consistently
\cite{black}.
Thus the net growth rate is proportional to 
$ |(\alpha_D+\alpha_F)k|$ for the mode $B_i({\bf
k},t)$. The expressions (\ref{divexp}) have apparent divergences at finite $q$; so in 
 perturbative calculations one should treat the $\alpha$-terms as
perturbations which remove these divergences. This problem is akin to that in
Kuramoto-Shivashinsky equation for flame front propagation \cite{proca}. So
the expressions for $\alpha_D$ and $\alpha_F$ are
\begin{eqnarray}
\alpha_D&=&{2S_3\over 3}\int {d^3q\over (2\pi)^3}{\tilde{D}_1(q)\over\nu (\nu+\mu)q^2},
\\
\alpha_F&=&-{2S_3\over 3}{4\over 15}\int {d^3q\over (2\pi)^3}{\tilde{D}_2(q,t)
\over 2\mu q^2},
\label{finexp}
\end{eqnarray}
which do not have any finite wavevector singularity. Expressions (\ref{finexp})
are obtained, as mentioned before, by truncating the perturbation series at the
tree level. Extensions to higher orders are straight forward. Illustrative
examples of higher order diagrams have been shown in Fig.\ref{oneloop}.

Let us now consider various $k$ dependences of $\tilde{D}_1(k)$ and
$\tilde{D}_2(k)$. When the background velocity field is driven by
the Navier-Stokes equation with a conserved noise (thermal noise) one requires
that $D_1(k)=D_1k^2,\,\tilde{D_1}=\tilde{D}_1|k|$, giving $\langle u_i({\bf
k},t)u_i({\bf -k},t)\rangle=constant$. If we assume similar $k$-dependences for
$\langle b_i({\bf k},0)b_i({-\bf k},0)\rangle$ then we require $D_2(k)\sim$
constant and $\tilde{D}_2(k)={\tilde{D}_2\over |k|}$. These choices yield
\begin{eqnarray}
\alpha_D&=&{2S_3\over 3}\int {d^3q\over
(2\pi)^3}{\tilde{D}_1|q|\over\nu(\nu+\mu)q^2},\nonumber
\\
\alpha_F&=&-{2S_3\over 3}{4\over 15}\int {d^3q\over
(2\pi)^3}{\exp[2(\alpha_D-\alpha_F)|q|t]\over
2\mu |q|},
\label{alphathermal}
\end{eqnarray}
which remain finite even if the system size diverges. 

A fully developed turbulent state, characterised by K41 energy spectra,  is
generated by $D_1(k)\sim k^{-3}$ and $\tilde{D}_1(k)=\tilde{D}_1k^{-4}$. In
addition if we assume that the initial magnetic fields correlation also have
K41 scaling then $D_2(k)\sim k^{-5/3}$ and
$\tilde{D}_2(k)=\tilde{D}_2k^{-8/3}$. If one starts with a K41-type initial
correlations for the magnetic fields, then at a later time the scale dependence
for the magnetic field correlations are likely to remain same; 
only the amplitudes grow.  Notice that the spectra diverge as wavevector
$k\rightarrow 0$, i.e.,  as the system size diverges. This is a typical
characteristic of fully developed turbulence. For such a system
we find
\begin{eqnarray}
\alpha_D&=&{2S_3\over 3}\int {d^3q\over (2\pi)^3}{\tilde{D}_1q^{-4}\over\nu
(\nu+\mu)q^2}={2S_3\over 3}\int {d^3q\over (2\pi)^3}{2M^{-1}D_1\over \nu^2 (1+P_m)
q^6},\nonumber \\
\alpha_F&=&-{2S_3\over 3}{4\over 15}\int {d^3q\over (2\pi)^3}{\tilde{D}_2(t)
q^{-8/3}\over 2\mu }=-{2S_3\over 3}{4\over 15}\int {d^3q\over
(2\pi)^3}{\tilde{D}_2(t)q^{-8/3}\over 2 P_m \nu }.
\label{alphak41}
\end{eqnarray}
The notable difference between the expressions Eqs.(\ref{alphathermal}) and
(\ref{alphak41}) for the $\alpha$ coefficients is that the $\alpha$
coefficients diverge with the system size if the energy spectra are singular
in the infra red limit (as in for fully developed turbulence). 
These divergences are
reminiscent of the divergences that appear in critical dynamics \cite{halparin} which are
handled by renormalisation group methods.

In general, at early times (small $\alpha_F$), $\alpha_F$ increases 
exponentially
in time. 
The growth rate of $\alpha_F$ decreases with time.
Since $\alpha_D$ and $\alpha_F$ have different signs, $
|(\alpha_D+\alpha_F)|\rightarrow 0$ as time $t$ increases. 
Thus the net growth rate comes down to zero. Hence, Eq.(\ref{alphathermal})
and Eq.(\ref{alphak41}) suggest that the early-time growth and late time
saturation of magnetic fields take place for different types of background
velocity correlations and initial magnetic field correlations. Therefore dynamo
instability and its saturation are rather intrinsic properties of the $3d$MHD
equations with broken reflection invariance. One may also note that for $K41$-type
of correlations (singular in the infrared limit) one has forward cascade of
kinetic energy \cite{yakhot}: This is because energy is fed into the system mostly
in the large scale (i.e., for small $k$) whereas, dissipation acts primarily in
the small scales (large $k$), resulting into a cascade of energy from the large-
to small- scales. On the other hand, for correlations smooth in the infra red
limit, there is no such cascade. These results indicate that 
the existence of the dynamo
mechanism does not require any special background velocity field spectrum,
though the value of the $\alpha$-coefficient depends upon it. 
Our results also suggest that these processes
may take place for varying magnetic Prandtl number $P_m=\mu/\nu$.
The above analysis crucially depends on the fact that $\alpha_F$ and 
$\alpha_D$ have opposite signs, which, in turn, imply that $\tilde{D}_1$
and $\tilde{D}_2$ have same signs. We have already seen
that in a physically realisable situation where parity is broken
entirely due to the global rotation,  $\tilde{D}_1$ and  $\tilde{D}_2$ indeed
have the same sign. 

In the
{\em first order smoothing approximation} \cite{moff,arnab} in the kinematic 
limit, to calculate 
$\langle \bf u\times b\rangle$ one considers only the Induction equation 
 as $\bf u$ is supposed to be given. However when one
goes beyond the kinematic approximation, one has to consider the Navier-Stokes
equation as well. Thus in the first-order smoothing approximation one writes
the equations for the fluctuations $\bf u$ and $\bf b$ as (to the first order)
\begin{equation}
{\partial {\bf b}\over\partial t}\approx \nabla \times ({\bf u\times 
\overline B}) + \nabla \times ({\bf \overline U \times b}),
\end{equation}
and
\begin{equation}
{\partial {\bf u}\over \partial t}\approx \ldots +({\bf \overline{B}.\nabla)
b},
\end{equation}
where the ellipsis refer to all other terms in the Navier-Stokes equation and 
$\bf \overline B$ and $\bf \overline U$ are the large scale ({\em mean field})
part of the velocity and magnetic fields \cite{moff,arnab}. With these
we can write
\begin{equation}
\langle {\bf u\times b}\rangle_i=\langle \epsilon_{ijp}u_jb_p\rangle=
\langle\epsilon_{ijp}u_jB_m
{\partial\over\partial x_m}u_p\rangle+\langle\epsilon_{ijp}b_pB_m
{\partial\over\partial x_m} b_j\rangle\equiv \alpha_{im}B_m+\ldots
\end{equation}
Here the ellipsis refer to non-$\alpha$ terms in the expansion of $\bf \langle
u\times b\rangle$ (see Eq.(4)).
Thus for isotropic situations $\alpha={\tau\over 3}[-\langle{\bf u.(\nabla\times
u)}\rangle+\langle{\bf b.(\nabla\times b)}\rangle]$ where $\tau$ is a
correlation time. 
Thus $\alpha$ is proportional to the difference in the fluid and 
magnetic {\em torsalities}\cite{po}, 
(fluid helicity being the same as fluid torsality and magnetic helicity being
 proportional to magnetic torsality)
a result obtained in \cite{po,field} using other methods and 
approximations. 
Note that  Eqs. (21) and 
(\ref{finexp}) are very similar to but not exactly the one that were
obtained in \cite{field} (in our notations $\tilde{D}_1$ is proportional to fluid
torsality (or fluid helicity) and $\tilde{D}_2$ is proportional to magnetic torsality). 
We ascribe this difference to the essential difference between a two-scale approach and
our dirgrammatic perturbation theory which, we believe is more suitable for handling
continuous kinetic and magnetic spectra. 

\section{Hydrodynamic limit of dynamo growth}
\label{renor}

We have seen that in Eqs.(\ref{alphak41}) the $\alpha$-coefficients diverge in the
hydrodynamic ($k\rightarrow 0$) limit which calls for a renormalisation group (RG) 
analysis as a natural extension of our diagrammatic perturbative calculations.
In fully developed $3d$MHD, in the steady state,
correlation and response functions exhibit dynamical scaling with the {\em
dynamic exponent} $z=2/3$ \cite{four,abjkb} (for a different approach
see \cite{verma}), which means renormalised
viscosities (kinetic as well as magnetic) diverge $\sim k^{-4/3}$ for a
wavenumber $k$ belonging to the inertial range. Even for decaying MHD with
initial K41-type correlations this turns out to be true \cite{decay} where
equal time correlations exhibit dynamical scaling with $z=2/3$. The question
is, what it is in the initial transient of dynamo growth ($ t\ll$ saturation
time). We examine this in a
renormalization group framework. Since we are interested in the early growth, we
neglect the Lorentz force and work with Eq.(\ref{indk}) inconjunction with the initial
magnetic field correlations and noise correlation given by
Eq.(\ref{varbi}). As before, we assume a statistical steady state for the velocity
field. It is well-known that correlations $\langle u_i({\bf k},t)u_j({\bf -k},0)\rangle$
exhibit scaling form $k^{-d-2\chi}h(tk^z)$ where $\chi$ and $z$ are the 
spatial scaling and 
dynamical exponents respectively \cite{yakhot} where $h$ is a scaling function. 
The Galilean invariance of the MHD equations constraints these exponents to obey
the relation $\chi+z=1$
\cite{yakhot,abjkb,abepl}. In addition to that, for fully developed turbulence
due to non-renormalization of the noise-correlators [cf. Eq.(\ref{varf})]
the exponents are fully
determined:
$z=2/3,\,\chi=1/3$, which means the renormalised fluid viscosity 
diverges as $k^{-4/3}$ 
in the limit wavevector $k\rightarrow 0$. During early growth, equal-time 
magnetic field correlations $\langle b_i({\bf k},t)b_j({\bf -k},t)\rangle$ are expected
to exhibit a scaling form $k^{-d-2\chi_b}m(tk^z_b)$ ($ t\ll$ saturation time)
where $\chi_b$ 
and $z_b$ are the magnetic spatial scaling
 and dynamical exponents respectively, and $m$ is a
scaling function. Similar conditions arising from the Galilean invariance and
non-renormalization of the initial K41-like magnetic field spectrum determines
$z=z_b=2/3$ and $\chi=\chi_b=1/3$.
We perform a renormalization
group analysis following \cite{fns,yakhot,decay}. As mentioned earlier, the
$\alpha$-term is treated as a perturbation. In a renormalisation-group
transformation, one integrates out a shell of modes $\Lambda e^{-l} <q <\Lambda$,
and and simultaneously rescales length scales, time intervals and fields through
 ${\bf x}\rightarrow e^l {\bf x},\, t\rightarrow e^{lz}t,\,{\bf u}\rightarrow e^{l\chi}{\bf
u},\, {\bf b}\rightarrow e^{l\chi}{\bf b}$. This has the effect that 
 the nonlinearities
are affected only by naive rescaling (this, a consequence of the Galilean
invariance of the $3d$MHD equations, essentially implies that the diagrammatic
corrections to the nonlinearties vanish in the long wavelength limit). 
The variances Eq.(\ref{varf}), which diverge at
low wavenumbers remain unrenormalised and thus affected only by rescaling. There
are however fluctuations corrections to $\mu$ and $\alpha_D$ which we evaluate at
the lowest order. The resulting RG 
flow equations for $\mu$ and $\alpha_D$,
obtained in a one-loop calculation are
\begin{eqnarray}
{d\mu\over dl}&=&\mu [z_b-2+ A_1{D_1\over \nu^2 (\nu+\mu)\Lambda^4}],\\
\label{flow1}
{d\alpha_D\over dl}&=&\alpha_D[z_b-1+A_2{\tilde{D}_1\over 
\alpha_D \nu (\nu+\mu)\Lambda^{3}}],
\label{flow2}
\end{eqnarray}
where $A_1,A_2$ are numerical constants. Equations (\ref{flow1}) and (\ref{flow2})
are similar to those presented in Ref.\cite{moff3} [Eqs. (10.13) and (10.14)] but
not exactly same. The differences arise mainly (apart from some detail 
technical differences in the perturbation theories involed)  from the fact
that in Ref.\cite{moff3} the expressions for the $\alpha$-coefficients were
derived for a given variance of the velocity field. In contrast, we use the
Navier-Stokes equation, driven by a stochastic force of given variance, in place
of a given velocity variance.
By substituting the value of the exponents in 
Eqs.(\ref{flow1}) and (\ref{flow2}) we find {\em renormalized} (i.e., wavevector
dependent) 
$\alpha_D (k)\sim \alpha_Dk^{-1/3},\,\mu (k)\sim k^{-4/3}$ in the hydrodynamic
($k\rightarrow 0$) limit. Thus in that limit, the effective dynamo equation 
takes the form
\begin{equation}
{\partial { b_i}\over \partial t}=(\alpha_D-\mu)k^{2/3}{ b_i}+....
\label{effeceq}
\end{equation}
where the ellipsis refer to non-linear terms and $i$ refers to the growing mode. 
Thus, in the hydrodynamic limit, 
there is growth of the magnetic fields only if $\alpha_D-\mu >0$. 
This can happen only if the
renormalised magnetic viscosity is
less than a critical value, set by $\alpha_D$, i.e., the kinetic helicity. 
In terms of the Ekman number $M$ this
condition means $M \stackrel{>}{\sim}1/2$ for anti dynamo, 
i.e., no growth, equivalently $M\lesssim 1/2$ for
growth of the magnetic fields. 
This can be achieved in two ways, namely by increasing rotation, keeping
the magnetic viscosity (or the magnetic Reynolds number) constant, 
or decreasing the magnetic
viscosity (i.e., increasing the magentic Reynolds number) for a constant 
rotation. This
conclusions are in good agreement with the numerical results of Ref.\cite{vin}.
Since renormalised
magnetic viscosity increases with its bare (microscopic) value, it suggests that bare
magnetic viscosity must be less than a critical value for growth to be possible. Thus
our RG results qualitatively explain the numerical results of Kida {\em et al} 
\cite{kida} who
they found that unless magnetic hyperviscosity was less than a critical 
value there was no
growth (it can be easily argued that a hypermagnetic viscosity gives rise to a
magnetic viscosity in the longer scale and hence their result in effect imposes a
critical value of the magnetic viscosity). In our model $\alpha$-effect is
proportional to $\tilde{D}$ which in turn is proportional to the global
rotation frequency. Hence our results suggest that $\alpha$-effect is likely to
grow with increasing rotational speed which is in agreement with the  results of
Ref.\cite{brand}. On the other hand, if the background velocity and the initial
magnetic field correlators do not have an infra red singularity (i.e., when the
correlators $\sim k^2$) there is no fluctuation correction to the magnetic
viscosity and to the $\alpha$-coefficient resulting in the fact that the
growth term ($\propto k$) dominates over the dissipation ($\propto k^2$) 
for sufficiently
small wavenumber $k$, leading to growth even for arbitrarily large magnetic
viscosity. Therefore, there is no critical $R_m$. 
Thus the effects of the infrared divergences that appear in the expressions for
the $\alpha$-coefficients [Eq. (\ref{alphak41})] are quite significant: They
indicate, as for the driven diffusive nonequilibrium systems with diverging
kinetic coefficients in the hydrodynamic limit \cite{fns,yakhot}, divergence of
time-scales in the hydrodynamic limit. Since, the $\alpha$-term in Eq.
(\ref{indk}) is proportional to wavenumber $k$, the time-scale of growth of the
mode with wavenumber $k$ is $O(\alpha k)$. This remains true, even in the
hydrodynamic limit, for the case when there is no divergence in the
$\alpha$-coefficients. In contrast, when the $\alpha$-coefficient diverge in the
infra red limit, the growth rate changes {\em qualitatively} from its linear
dependence on wavenumber $k$ in the hydrodynamic limit. For example, with the
the background velocity correlations and the 
initial magnetic field correlations given by
Eq. (\ref{varf}), the $\alpha$ coefficients diverge as $k^{-1/3}$ in the long
wavelength limit. Hence, the effective growth rate 
is changed to $\alpha (k)k\sim k^{2/3}$.
A full self-consistent calculation (when  feedback due to the Lorentz force
cannot be neglected) for the
$\alpha$-coefficients require simultaneous solutions of the self-consistent expressions
for magnetic Prandtl number, magnetic- to kinetic- energy ratio and the
$\alpha$-coefficients which can be handled in our scheme of calculations. The
self-consistent solutions are expected to be influenced by the degree of
crosscorrelations between the velocity and magnetic fields \cite{abepl}.

So far, we have assumed that both $D_1(k)$ and $\tilde{D}_1(k)$ have the same
infra red singularity ($D_1(k)\sim k^{-5/3}$ and $\tilde{D}_1(k)\sim k^{-5/3}$).
This need not be the case always. However, if $\tilde{D}_1(k)$ is non-singular
then $\alpha_D$ does not diverge. As a result, the growth rate is just $\alpha_D
k$ even in the hydrodynamic (long wavelength) limit. Effective dissipation,
however, will still be $\sim k^{2/3}$ and thus it will dominate over $O(k)$
growth. Therefore, there will be no growth in the hydrodynamic limit. Thus, our
analyses suggest that in any fully developed turbulent system  with
$\alpha$-effect, helicity spectrum (given by $\tilde{D}_1(k)$) 
should be as singular as the kinetic energy spectrum (given by $D_1(k)$).

\section{conclusions}
\label{summ}
In conclusions, we have calculated expressions for the $\alpha$-coefficients in
a diagrammatic perturbation theory on a minimal model
 for arbitrary background velocity and initial magnetic
field correlations, and fluid and magnetic viscosities.
We show that the parity
breaking parts of the velocity and magnetic field variances must have the same
sign, which is the case in any physical system. We explicitly show that
the processes of early growth and late-time saturations may take place
independent of any special velocity and initial magnetic field correlations.
Even though our explicit
calculations were done by using simple initial conditions for the
calculational convenience, the results that we obtain are general
enough and it is apparent that the feedback mechanism is qualitatively 
independent
of the details of the initial conditions and force correlations. one may note
that for one of the force/initial correlations there is no kinetic energy 
cascade in the conventional sense
but we still find dynamo action. It is quite reasonable to expect that
our results should be 
valid for more realistic initial conditions also. In effect we have explicitly
demonstrated the robustness and generality of the dynamo mechanism and that
the dynamo mechanism is an intrinsic property of the $3d$MHD equations. 
We have also shown, within our RG
analysis, that the magnetic viscosity should be less than a critical value for 
growth of magnetic fields a result which was previouly observed in
numerical simulations. We conclude the existence of a critical Ekman number
for K41-type correlations: We find growth only when $M\lesssim 1/2$, 
confirming recent
numerical results. This is easily understood
in our framework.
The issue of divergent effective viscosities in the inertial
range assumes importance as it may help to overcome some of the non-linear
restrictions as discussed by Vainshtein and Cattaneo \cite{vain}. 
A system of magnetohydrodynamic turbulence in a rotating
frame, after the saturation time (i.e., after which there is no net 
growth of the magnetic fields) belongs to the universality class of usual three-dimensional 
magnetohydrodynamic turbulence in a laboratory. This can be seen easily 
in both the lab and the rotating frames; the critical exponents characterising the
correlation functions
can be calculated exactly by using the Galilean invariance and
noise-nonrenormalisation conditions \cite{yakhot,abjkb}. An important question,
which remains open for further investigations, is the multiscaling properties of
the velocity and the magnetic field structure functions at various stages of the
growth of the magnetic fields. In what concerns an experimental observation of our
results, one should add that even though it is not  easy
 to verify our results in an experimental set up,
numerical simulations of Eqs.(\ref{nsk}) and (\ref{indk}) with the variances
(\ref{varbi}) with different $k$-dependences can be performed to check
these results.

\section{Acknowledgement}
The author wishes to thank J. K. Bhattacharjee for drawing his attention to
this problem, and R. Pandit, J. Santos  and the anonymous referee
for many fruitful comments and suggestions.
The author thanks the Alexander von Humboldt Foundation, Germany for 
financial support.

\end{document}